\documentclass[aps,pra,preprintnumbers,showpacs,tightenlines]{revtex4}

\usepackage{amssymb}
\usepackage{amsmath}
\usepackage{graphicx}
\usepackage{epsfig}
\usepackage{subfigure}
\usepackage{amsfonts}
\usepackage{CJK}

\begin{document}

\title{Simultaneous quantum state exchange or transfer between two sets of cavities and
generation of multiple Einstein-Podolsky-Rosen pairs via a superconducting coupler qubit}

\author{Chui-Ping Yang$^{1}$, Qi-Ping Su$^{1}$, Shi-Biao Zheng$^{2}$, and Siyuan Han$^{3}$}

\address{$^1$Department of Physics, Hangzhou Normal University,
Hangzhou, Zhejiang 310036, China}

\address{$^2$Department of Physics, Fuzhou University, Fuzhou 350002, China}

\address{$^3$Department of Physics and Astronomy, University of
Kansas, Lawrence, Kansas 66045, USA}

\date{\today}

\begin{abstract}
We propose an approach to simultaneously perform quantum state exchange or
transfer between two sets of cavities, each containing $N$ cavities, by using only one superconducting
coupler qubit. The quantum states to be exchanged or transferred can be arbitrary
pure or mixed states and entangled or nonentangled. The operation time does not increase with the number
of cavities, and there is no need of applying classic pulses during the
entire operation. Moreover, the approach can be also applied to realize
quantum state exchange or transfer between two sets of qubits, such as that
between two multi-qubit quantum registers. We further show that the present proposal
can be used to simultaneously generate multiple Einstein-Podolsky-Rosen pairs of photons or qubits
, which are important in quantum communication. The method can be generalized to
other systems by using different types of physical qubit as a coupler to
accomplish the same task.
\end{abstract}

\pacs{03.67.-a, 42.50.Pq, 85.25.-j} \maketitle
\date{\today}

\begin{center}
\textbf{I. INTRODUCTION}
\end{center}

Physical systems composed of cavities (a.k.a. resonators) and
superconducting qubits such as charge, flux, transmon and phase qubits are
considered as one of the most promising candidates for quantum information
processing (QIP) [1]. Reasons for this are as follows. First, the use of
conventional microfabrication techniques allows straightforward scaling to
large numbers of stationary qubits. Second, circuit cavities can be used to
engineer a variety of qubit types and interactions [2-6]. Last, the strong
coupling limit can readily be achieved with superconducting qubits coupled
to circuit cavities, which was earlier predicted theoretically [4,7] and has
been experimentally demonstrated [8,9].

During the past decade, much progress has been made in quantum state
engineering and quantum logic operations with superconducting qubits coupled
to a single superconducting cavity. For instance, theoretical proposals have
been presented for generating various quantum states of a single
superconucting cavity (e.g., Fock states, coherent states, squeezed states,
the Sch\"ordinger Cat state, and an arbitrary superposition of Fock states)
[10-12]; and experimental preparation of a Fock state and a superposition of
Fock states of a superconducting cavity has been reported [13,14]. On the
other hand, many schemes have been proposed for realizing quantum logical
gates and generating quantum entanglement with two or more superconducting
qubits coupled to a cavity (usually in the form of coplanar waveguide
resonator) [2-4,7,14-19]. So far, quantum logic operations [6, 20-23] and
entanglement [20,25-28] involving two [6,20] or three qubits [21-23] have
been experimentally demonstrated with superconducting qubits coupled to a
single cavity. Furthermore, quantum information transfer between two qubits
[6], three-qubit quantum error correction [22], two-qubit Grover search and
Deutsch-Jozsa quantum algorithms [24], and three-qubit Shor's algorithm to
factor the number 15 [28] have also been implemented in such a system.

Attention is now shifting to quantum state engineering with larger systems
composed of multiple cavities (and qubits) instead of a single cavity.
Within the circuit QED architecture, several theoretical schemes for
generating entangled photon Fock states of two resonators have been proposed
[29,30]. By coupling two resonators via a phase qubit, Wang et al. recently
demonstrated an entangled NOON state of photons in two superconducting
microwave resonators [31]. Moreover, proposals for generating entanglement
of photons and superconducting qubits in multiple cavities coupled by a
superconducting qutrit have been presented recently [32,33].

In this paper, we focus on another interesting aspect, i.e., quantum state
exchange and transfer among multiple cavities. The former is necessary to
perform QIP involving qubits (photons or other matter qubits)\ located in
different cavities, and the latter is needed for efficient transfer of
quantum information from solid state quantum registers specialized in
processing quantum information to quantum memory cells (consisting of long
lifetime qubits) for long time storage. Besides their significance in QIP,
it is also known that both quantum state exchange and transfer are important
in quantum (networked) communication. In addition, we will consider
generation of multiple Einstein-Podolsky-Rosen (EPR) pairs of photons or
qubits, which are important in quantum communication.

In the following, we propose an approach to exchange quantum states or
transfer between two sets of cavities [Fig~1(a)], each containing $N$
cavities, by using only \textit{one} superconducting coupler qubit. The
approach can be also applied to simultaneously generating multiple EPR pairs
of photons, as shown below. Our proposal has the following advantages: (i)
the procedure for implementing quantum state exchange or transfer is
independent of the initial states of cavities (either pure or mixed states
and entangled or nonentangled), (ii) quantum state exchange or transfer
between different pairs of cavities can be simultaneously performed, and
multiple EPR pairs of photons or qubits can be created at the same time,
(iii) the operation time does not increase with the number of cavities
involved, (iv) no classical microwave pulse is needed for the operation, and
(v) only one superconducting qubit is used as a coupler.

To the best of our knowledge, no scheme has been reported for implementing
quantum state exchange or transfer between two sets of cavities, and no
proposal has been presented for simultaneously generating multiple EPR pairs
of photons or qubits distributed in different cavities, and our proposal is
the first one to demonstrate that all these tasks can be implemented in
circuit QED by using only \textit{one} superconducting qubit as a coupler.
Because only one superconducting coupler qubit is needed, the circuit is
greatly simplified and the engineering complexity and cost is much reduced.
The method presented here is quite general and can be applied to other
systems by using different types of physical qubit (e.g., a quantum dot) as
a coupler to accomplish the same task.

Furthermore, the present method can be applied to exchanging or transferring
quantum states between two sets of qubits (e.g., two multi-qubit quantum
registers). To see this, let us look at Fig.~1(b), which shows a set of $N$
qubits distributed in one set of $N$ cavities ($a_1,a_2,...,a_N$), and the
other set of $N $ qubits located in another set of cavities ($%
b_1,b_2,...,b_N $), respectively. The qubits embedded in cavities can be
solid-state qubits, atoms, etc. Exchanging or tranferring quantum states
between the two sets of qubits can be done as follows. First, by performing
local operations (i.e., each local operation is performed on one qubit in
each cavity such that the state of the qubit in each cavity is transferred
onto the field of each cavity), one can map the state of each qubit to the
cavity in which it is located. Second, one can use the present method to
exchange or transfer quantum states between these two sets of cavities.
Last, map the state of each cavity back to the respective qubit through
local operations. In this way, quantum states can be exchanged or
transferred between these two sets of qubits. In a similar manner, $N$ EPR
pairs of photons simultaneously generated can be transferred onto qubits
distributed in $2N$ cavities, by performing local operations within each
cavity.

\begin{figure}[tbp]
\includegraphics[bb=62 384 551 614, width=10.5 cm, clip]{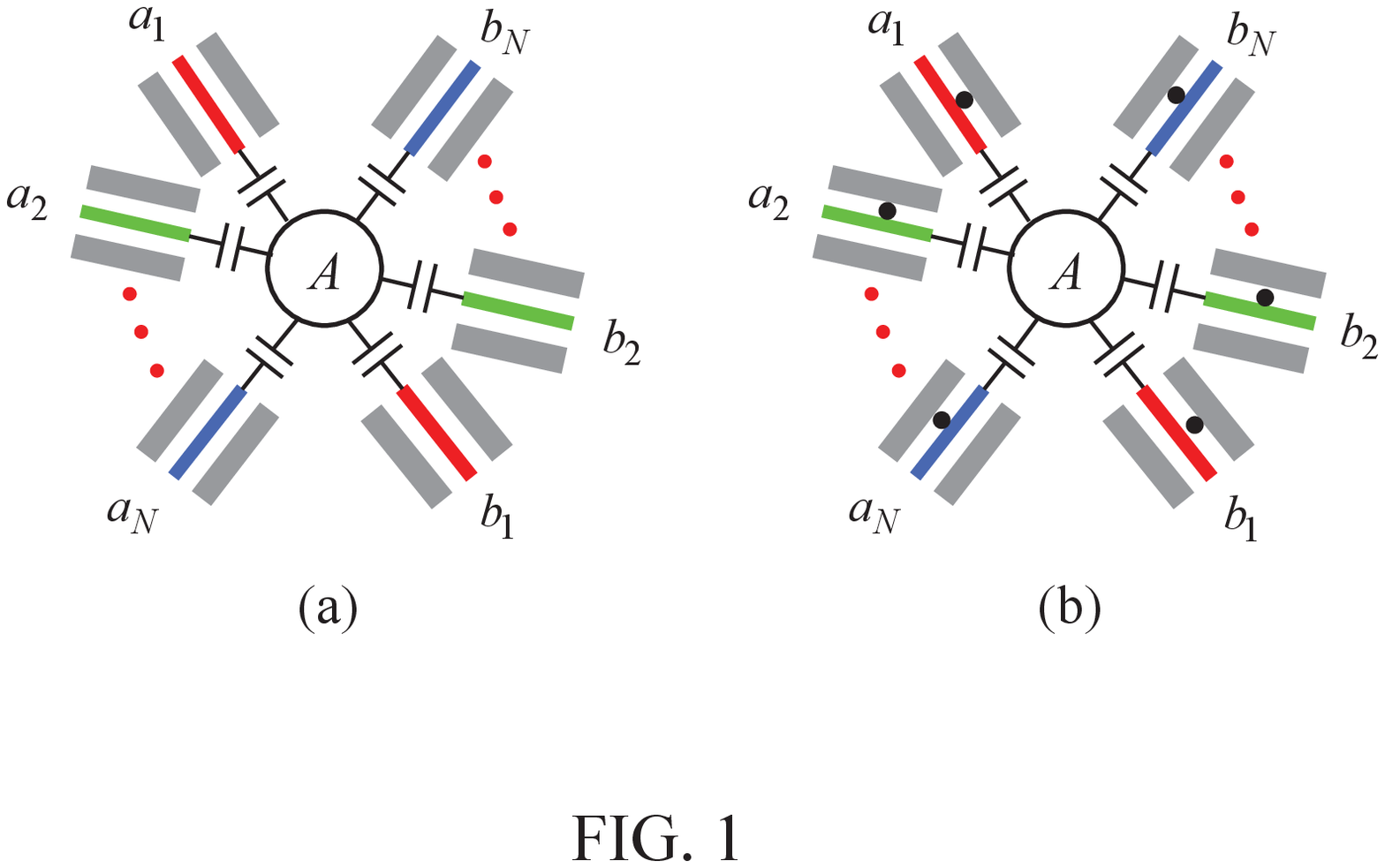} \vspace*{%
-0.08in}
\caption{(Color online) (a) Diagram of a superconducting qubit $A$ (a circle
at the center) and $2N$ cavities. The first set of $N$ cavities are labelled
as $a_1,a_2,...,$ and $a_N$, while the second set of $N$ cavities are
denoted as $b_1,b_2,...,$ and $b_N$. Each cavity could be a one-dimensional
coplanar waveguide resonator which is capacitively coupled to the coupler
qubit $A$. (b) The diagram of a superconducting qubit $A$ and $2N$ cavites ,
each cavity hosting a qubit (labelled by a dark dot), which can be a
solid-state qubit, an atom, etc. Quantum states can be exchanged or
transferred between $N$ qubits (in one set of $N$ cavities) and another $N$
qubits (in the other set of $N$ cavities). In addition, $N$ EPR pairs can be
simultaneously generated using photons or qubits distributed in $2N$
cavities. For details, see the discussion given in the text.}
\label{fig:1}
\end{figure}

The paper is organized as follows. In Sec.~II, we show how to exchange
quantum states between two sets of cavities by using a superconducting
coupler qubit interacting with these cavities. Under the large detuning
condition, the coupler qubit does not exchange energy with the cavities, but
it can mediate coupling between any pair of cavities whose quantum states
are to be exchanged. We further discuss the fidelity of operation by
considering the factors that may reduce fidelity, and then estimate the
fidelity numerically for several initial states of four cavities coupled to
a phase qubit as an example. In Sec.~III, we show how to simulateously 
generate multiple EPR pairs of photons or qubits by using the present proposal.
A concluding summary is given in Sec.~IV.

\begin{center}
\textbf{II. MULTI-CAVITY QUANTUM STATE EXCHANGE AND TRANSFER}
\end{center}

Consider $2N$ cavities coupled to a superconducting qubit $A$ [Fig.~1(a)].
The first set of $N$ cavities are labeled as cavities $a_1,a_2,...,$ and $%
a_N $ while the second set of $N$ cavities are labeled as cavities $%
b_1,b_2,...,$and $b_N$. In addition, the two levels of the qubit $A$ are
denoted as $\left| g\right\rangle $ and $\left| e\right\rangle $ (Fig.~2).
Suppose that cavity $a_j$ ($b_j$) with $j=1,2,...,N$ is coupled to the $%
\left| g\right\rangle $ $\leftrightarrow $ $\left| e\right\rangle $
transition with coupling strength $g_j$ ($\mu _j$) and detuning $\Delta
_j=\omega _{eg}-\omega _{aj}=\omega _{eg}-\omega _{bj}$ (Fig.~2)$.$ Here, $%
\omega _{aj} $ ($\omega _{bj}$) is the frequency of cavity $a_j$ ($b_j$). In
the interaction picture, the Hamiltonian of the whole system is given by
\begin{equation}
H=\sum_{j=1}^N\left( g_je^{i\Delta _jt}\hat a_jS_{+}+\mu _je^{i\Delta
_jt}\hat b_jS_{+}+\text{H.c.}\right) ,
\end{equation}
where $S_{+}=\left| e\right\rangle \left\langle g\right| $, and $\hat a_j$ ($%
\hat b_j$) is the photon annihilation operator of cavity $a_j$ ($b_j$).

\begin{figure}[tbp]
\includegraphics[bb=116 312 455 668, width=6.5 cm, clip]{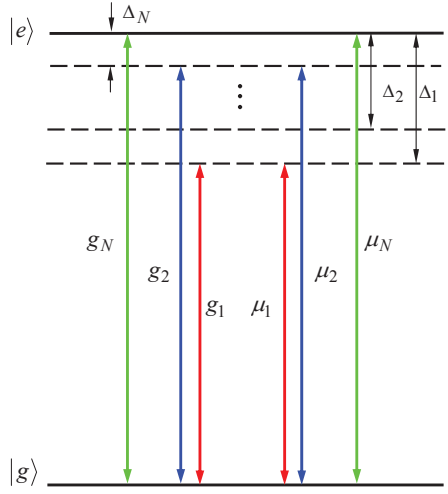} \vspace*{%
-0.08in}
\caption{(Color online) Illustration of qubit-cavity dispersive interaction.
The mode of cavity $a_j$ (cavity $b_j$) is dispersively coupled to the $%
\left| g\right\rangle \leftrightarrow \left| e\right\rangle $ transition
with a coupling constant $g_j$ ($\protect\mu _j$) and detuning $\Delta_j$ ($%
j=1,2,...,N$).}
\label{fig:2}
\end{figure}

Under the large detuning condition $\Delta _j\gg g_j,\mu _j,$\ the cavities
do not exchange energy with the qubit. However, the qubit can mediate
dispersive coupling between the cavities. Cavity $a_j$\ is coupled to $b_j$\
only when the detunings satisfy the following conditions
\begin{equation}
\frac{\left| \Delta _j-\Delta _k\right| }{\Delta _j^{-1}+\Delta _k^{-1}}\gg
g_jg_k,\;g_j\mu _k,\;\mu _j\mu _k;\;\text{ }j\neq k.
\end{equation}
Then we can obtain the following effective Hamiltonian

\begin{equation}
H_{eff}=H_0+H_I,
\end{equation}
with
\begin{eqnarray*}
H_0 &=&\sum_{j=1}^N\left( \frac{g_j^2}{\Delta _j}\hat a_j\hat a_j^{+}+\frac{%
\mu _j^2}{\Delta _j}\hat b_j\hat b_j^{+}\right) \left| e\right\rangle
\left\langle e\right| \\
&&-\sum_{j=1}^N\left( \frac{g_j^2}{\Delta _j}\hat a_j^{+}\hat a_j+\frac{\mu
_j^2}{\Delta _j}\hat b_j^{+}\hat b_j\right) \left| g\right\rangle
\left\langle g\right| ,
\end{eqnarray*}
\begin{equation}
H_I=\sum_{j=1}^N\lambda _j(\hat a_j\hat b_j^{+}+\hat a_j^{+}\hat b_j)(\left|
e\right\rangle \left\langle e\right| -\left| g\right\rangle \left\langle
g\right| ),
\end{equation}
where $\lambda _j=g_j\mu _j/\Delta _j,$\ and the two terms in the first
(second) bracket of $H_0$\ are ac-Stark shifts of the level $\left|
e\right\rangle $\ ($\left| g\right\rangle $) induced by the cavity modes $%
a_j $\ and $b_j$, respectively.

In the following, we set $g_j=\mu _j$ (achievable by tuning the coupling
capacitance between the qubit and cavity $a_j$ as well as the coupling
capacitance between the qubit and cavity $b_j$). In a new interaction
picture under the Hamiltonian $H_0,$ we have
\begin{equation}
\widetilde{H}_I=e^{iH_0t}H_Ie^{-iH_0t}=H_I.
\end{equation}
When the qubit is initially in the lower level $\left| g\right\rangle $, it
will remain in this state throughout the interaction. Thus, based on Eqs.
(4) and (5), one can see that the effective Hamiltonian goverening the field
dynamics is then given by

\begin{equation}
H_{e}=-\sum_{j=1}^{N}\lambda _{j}(\hat{a}_{j}\hat{b}_{j}^{+}+\hat{a}_{j}^{+}%
\hat{b}_{j}),
\end{equation}%
which leads to the tranformations
\begin{eqnarray}
e^{-iH_{e}t}\hat{a}_{j}^{+}e^{iH_{e}t} &=&\cos (\lambda _{j}t)\hat{a}%
_{j}^{+}+i\sin (\lambda _{j}t)\hat{b}_{j}^{+},  \notag \\
e^{-iH_{e}t}\hat{b}_{j}^{+}e^{iH_{e}t} &=&\cos (\lambda _{j}t)\hat{b}%
_{j}^{+}+i\sin (\lambda _{j})\hat{a}_{j}^{+}.
\end{eqnarray}%
In the following, we set $\lambda _{j}=\lambda ,$ i.e., $g_{j}\mu
_{j}/\Delta _{j}=\lambda $ (independent of $j$), which can be met by
adjusting the frequencies of cavities $a_{j}$ and $b_{j}$ such that $\Delta
_{j}=g_{j}\mu _{j}/\lambda .$ For $\lambda t=\pi /2,$ we obtain the
following transforms
\begin{eqnarray}
e^{-iH_{e}t}\hat{a}_{j}^{+}e^{iH_{e}t} &=&i\hat{b}_{j}^{+},\text{ \ }%
e^{-iH_{e}t}\hat{a}_{j}e^{iH_{e}t}=-i\hat{b}_{j},  \notag \\
e^{-iH_{e}t}\hat{b}_{j}^{+}e^{iH_{e}t} &=&i\hat{a}_{j}^{+},\text{ \ }%
e^{-iH_{e}t}\hat{b}_{j}e^{iH_{e}t}=-i\hat{a}_{j}.
\end{eqnarray}

Any initially unentangled field state of the first set of $N$ cavities $%
(a_{1},a_{2},...,a_{N})$ and the second set of $N$ cavities ($%
b_{1},b_{2},...,b_{N}$) can be described by the density operator $\rho
^{a}\left( 0\right) \otimes \rho ^{b}\left( 0\right) ,$ where the first
(second)\ part of the product is initial density operator of the first
(second) set of $N$ cavities, taking a general form of
\begin{eqnarray}
\rho ^{a}\left( 0\right) &=&\sum_{n_{j},m_{j}=0}^{\infty
}P_{\{n_{j},m_{j}\}}\prod_{j=1}^{N}\left\vert n_{j}\right\rangle
_{a_{j}}\left\langle m_{j}\right\vert , \\
\rho ^{b}\left( 0\right) &=&\sum_{s_{k},t_{k}=0}^{\infty
}P_{\{s_{k},t_{k}\}}\prod_{k=1}^{N}\left\vert s_{k}\right\rangle
_{b_{k}}\left\langle t_{k}\right\vert ,
\end{eqnarray}%
where the subscript $a_{j}$\ ($b_{k}$) represents cavity $a_{j}$\ ($b_{k}$)
as mentioned above, $P_{\{n_{j},m_{j}\}}$\ is the coefficient of the
component $\prod_{j=1}^{N}\left\vert n_{j}\right\rangle _{a_{j}}\left\langle
m_{j}\right\vert $\ of the initial density operator for the cavities ($%
a_{1},a_{2},...,a_{N}$), and the same notation applies to $%
P_{\{s_{k},t_{k}\}}$\ for the cavities ($b_{1},b_{2},...,b_{N}$). In terms
of $\left\vert n_{j}\right\rangle _{a_{j}}=\frac{\hat{a}_{j}}{\sqrt{n_{j}!}}%
\left\vert 0\right\rangle _{a_{j}}$ and $\left\vert s_{k}\right\rangle
_{b_{k}}=\frac{\hat{b}_{k}^{+s_{k}}}{\sqrt{s_{k}!}}\left\vert 0\right\rangle
_{b_{k}},$ we can write down the initial state as
\begin{eqnarray}
&&\ \ \ \rho ^{a}\left( 0\right) \otimes \rho ^{b}\left( 0\right)  \notag \\
\ &=&\sum_{n_{j},m_{j}=0}^{\infty
}P_{\{n_{j},m_{j}\}}\sum_{s_{k},t_{k}=0}^{\infty
}P_{\{s_{k},t_{k}\}}\prod_{j=1}^{N}\prod_{k=1}^{N}  \notag \\
&&\left( \frac{\hat{a}_{j}^{+n_{j}}\hat{b}_{k}^{+s_{k}}}{\sqrt{n_{j}!s_{k}!}}%
\left\vert 0\right\rangle _{a}\left\langle 0\right\vert \otimes \left\vert
0\right\rangle _{b}\left\langle 0\right\vert \frac{\hat{a}_{j}^{m_{j}}\hat{b}%
_{k}^{t_{k}}}{\sqrt{m_{j}!t_{k}!}}\right) .
\end{eqnarray}%
where $\left\vert 0\right\rangle _{a}=\left\vert 0\right\rangle
_{a_{1}}...\left\vert 0\right\rangle _{a_{N}}$ and $\left\vert
0\right\rangle _{b}=\left\vert 0\right\rangle _{b_{1}}...\left\vert
0\right\rangle _{b_{N}}.$

Based on Eq. (11), one can easily find that under the Hamiltonian $H_{e},$
the state of the cavity system after an evolution time $t=\pi /\left(
2\lambda \right) $ is given by
\begin{eqnarray}
\rho _{c}\left( t\right) &=&e^{-iH_{e}t}\rho ^{a}\left( 0\right) \rho
^{b}\left( 0\right) e^{iH_{e}t}  \notag \\
\ &=&\sum_{n_{j},m_{j}=0}^{\infty }\sum_{s_{k},t_{k}=0}^{\infty
}P_{\{n_{j},m_{j}\}}P_{\{s_{k},t_{k}\}}\prod_{j=1}^{N}\prod_{k=1}^{N}  \notag
\\
&&\left[ \frac{\left( i\hat{b}_{j}^{+}\right) ^{n_{j}}\left( i\hat{a}%
_{k}^{+}\right) ^{s_{k}}}{\sqrt{n_{j}!s_{k}!}}\left\vert 0\right\rangle
_{a}\left\langle 0\right\vert \right.  \notag \\
&&\left. \otimes \left\vert 0\right\rangle _{b}\left\langle 0\right\vert
\frac{\left( -i\hat{b}_{j}\right) ^{m_{j}}\left( -i\hat{a}_{j}\right)
^{t_{k}}}{\sqrt{m_{j}!t_{k}!}}\right]  \notag \\
\ &=&\sum_{s_{k},t_{k}=0}^{\infty }P_{\{s_{k},t_{k}\}}\prod_{k=1}^{N}\left[
i^{s_{k}}\left( -i\right) ^{t_{k}}\left\vert s_{k}\right\rangle
_{a_{k}}\left\langle t_{k}\right\vert \right]  \notag \\
&&\otimes \sum_{n_{j},m_{j}=0}^{\infty }P_{\{n_{j},m_{j}\}}\prod_{j=1}^{N}
\left[ i^{n_{j}}\left( -i\right) ^{m_{j}}\left\vert n_{j}\right\rangle
_{b_{j}}\left\langle m_{j}\right\vert \right] .  \notag \\
&&
\end{eqnarray}%
where we have used the unitary transformations described by Eq. (8). Note
that in the last two lines of Eq. (12), the first part of the product
represents the $N$-cavity state of ($a_{1},a_{2},...,a_{N}$) while the
second part is that of ($b_{1},b_{2},...,b_{N}$).

After returning to the original interaction picture, the state of the whole
system is given by
\begin{equation}
\rho _{cA}^{\prime }\left( t\right) =e^{-iH_0t}\rho _c\left( t\right) \rho
_A\left( t\right) e^{iH_0t}=\rho _c^{\prime }\left( t\right) \otimes \rho
_A\left( t\right) ,\;\;
\end{equation}
where $\rho _A\left( t\right) =\rho _A\left( 0\right) =\left| g\right\rangle
\left\langle g\right| ,$ and
\begin{equation}
\rho _c^{\prime }\left( t\right) =\rho ^a\left( t\right) \otimes \rho
^b\left( t\right)
\end{equation}
with
\begin{equation}
\rho ^a\left( t\right) =\sum_{s_k,t_k=0}^\infty
P_{\{s_k,t_k\}}^a\prod_{k=1}^N\left[ e^{i\phi _k(s_k-t_k)\pi }\left|
s_k\right\rangle _{a_k}\left\langle t_k\right| \right] ,
\end{equation}
\begin{equation}
\rho ^b\left( t\right) =\sum_{n_j,m_j=0}^\infty
P_{\{n_j,m_j\}}^b\prod_{j=1}^N\left[ e^{i\theta _j(n_j-m_j)\pi }\left|
n_j\right\rangle _{b_j}\left\langle m_j\right| \right] ,
\end{equation}
where $\phi _k=1/2+\left( g_k^2/\Delta _k\right) /(2\lambda )$ and $\theta
_j=1/2+\left( \mu _j^2/\Delta _j\right) /(2\lambda )$. This is equivalent to
the quantum state swap operation plus additional photon-number-dependent
phase shifts on the respective cavities. For implementation of a quantum
information processing task, these phase shifts can be absorbed into the
corresponding local operations.

In the above, we have shown how to implement quantum state exchange between
one set of $N$ cavities ($a_1,a_2,...,a_N$) and another set of $N$ cavities (%
$b_1,b_2,...,b_N$). It should be mentioned here\textbf{\ }that when each of
the second set of cavities is initially in the vacuum state, the quantum
state exchange protocol described above becomes quantum state transfer from
the first to the second set of cavities.

The method presented above can in principle be used to implement $N$-cavity
arbitrary quantum state exchange and transfer. However, as the number of
cavities increases it may become more difficult to satisfy condition (2). In
addition, beyond a certain number the coupling strength of the coupler qubt $%
A$\ with each cavity decreases as the number of cavities increases, which
leads to a longer operation time. Thus, decoherence, caused by relaxation
and dephasing of qubit and/or cavity decay, may become a bigger problem.
Nevertheless, we remark that when\ the number of cavities coupled to qubit $%
A $\ is limited to about 4 to 6, condition (2) can be readily satisfied and
the qubit-cavity couplings can remain sufficiently strong.

We should mention that the Hamiltonian (6) with $j=1$, i.e., $\lambda (\hat{a%
}\hat{b}^{+}+\hat{a}^{+}\hat{b}),$ was previously used in quantum conversion
between two cavity fields (or two coupled resonators), quantum conversion
between light and a macroscopic oscillator, quantum conversion between light
and the motion of a trapped atom (or the center-of-mass motion of an ion),
and quantum state transfer from light to matter [34-40]. However, it is
worth noting that the main purpose of this work is to construct the
Hamiltonian (6) with $j>1,$ the main result of this paper, which is
nontrival because it can be used to achieve simultaneous quantum state
exchange or transfer among multiple cavities, as shown above.

Furthermore, the present proposal can be used to transfer a \textit{%
multi-cavity entangled state} from $N$ cavities to another $N$ cavities.
This result is remarkable when compared with the previous proposals [34-40].
It is noted that the latter can only be used to transfer a single-cavity
quantum state (\textit{not entangled}) from one cavity to another cavity,
because they are based on the Hamiltonian $\lambda (\hat{a}\hat{b}^{+}+\hat{a%
}^{+}\hat{b})$ only for two cavities.

Finally, we should point out that a Hamiltonian $\sum_{j=1}^{N}\lambda _{j}(%
\hat{a}_{j}\hat{a}_{j+1}^{+}+\hat{a}_{j}^{+}\hat{a}_{j+1})$, describing the
interaction between two neighbor cavities in an array of cavities, has been
discussed previously (e.g., see [41,42]). It is obvious that this
Hamiltonian is different from ours given in Eq. (6).

We stress that this work is of interest in at least two aspects. First, the
protocol can be used to swap or transfer any type of multipartite
entanglement (e.g., the GHZ state $\left\vert 00...0\right\rangle
+\left\vert 11...1\right\rangle ,$ the W state $\frac{1}{\sqrt{N}}\left(
\left\vert 00...001\right\rangle +\left\vert 00...010\right\rangle
+...+\left\vert 10...000\right\rangle \right) ,$ the cluster state, etc.)
from one set of $N$ cavities to another set of $N$ cavities via a single
intermediate coupler qubit only. Second, the protocol can be used to
simultaneously generate multiple EPR pairs of photons or qubits, as shown
below.

As an example, let us consider four cavities coupled via a superconducting
qubit. In the following, we will first give a general discussion on the
fidelity of the operation. To quantify how well the proposed protocol works
out, we then estimate the fidelity numerically for several initial states of
the four-cavity ($N=2$) system.

The dynamics of the system, with finite qubit relaxation and dephasing and
photon lifetime included,, is determined by

\begin{eqnarray}
\frac{d\rho }{dt} &=&-i\left[ H,\rho \right] +\sum_{j=1}^{2}\kappa _{j}%
\mathcal{L}\left[ \hat{a}_{j}\right] +\sum_{j=1}^{2}\kappa _{j}^{\prime }%
\mathcal{L}\left[ \hat{b}_{j}\right]  \notag \\
&&+\gamma _{\varphi }\left( S_{z}\rho S_{z}-\rho \right) +\gamma \mathcal{L}%
\left[ S_{-}\right] ,
\end{eqnarray}%
where $H$ is Hamiltonian (1), $\mathcal{L}\left[ \hat{a}_{j}\right] =\hat{a}%
_{j}\rho \hat{a}_{j}^{+}-\hat{a}_{j}^{+}\hat{a}_{j}\rho /2-\rho \hat{a}%
_{j}^{+}\hat{a}_{j}/2,$ $\mathcal{L}\left[ \hat{b}_{j}\right] =\hat{b}%
_{j}\rho \hat{b}_{j}^{+}-\hat{b}_{j}^{+}\hat{b}_{j}\rho /2-\rho \hat{b}%
_{j}^{+}\hat{b}_{j}/2,$ and $\mathcal{L}\left[ S_{-}\right] =S_{-}\rho
S_{+}-S_{+}S_{-}\rho /2-\rho S_{+}S_{-}/2$. In addition, $\kappa _{j}$ ($%
\kappa _{j}^{\prime }$) is the photon decay rate of cavity $a_{j}$ ($b_{j}$)$%
,$ $\gamma _{\varphi }$ and $\gamma $ are the dephasing rate and the energy
relaxation rate of the level $\left\vert e\right\rangle $ of the qubit,
respectively.

The fidelity of the operation is given by
\begin{equation}
\mathcal{F}=T_{r}\sqrt{\rho _{id}\widetilde{\rho }^{1/2}\rho _{id}},
\end{equation}%
where $\rho _{id}$ is the state of an ideal system (i.e., without
relaxation, dephasing, photon decay, crosstalks, etc.) and $\widetilde{\rho }
$ is the final density operator of the system when the operation is
performed in a realistic physical system. From the description given in the
previous section, one can see that
\begin{equation}
\rho _{id}=\rho ^{a}(t)\rho ^{b}(t)\otimes \left\vert g\right\rangle
\left\langle g\right\vert ,
\end{equation}%
where $\rho ^{a}(t)$\ and $\rho ^{b}(t)$\ are, respectively, the states of
cavities ($a_{1},a_{2}$) and ($b_{1},b_{2}$) given in Eqs. (15) and (16) for%
\textbf{\ }$N=2$\textbf{.}

According to the previous section, we have $g_1=\mu _1,$ $g_2=\mu _2,$ $g_2=%
\sqrt{\frac{\Delta _2}{\Delta _1}}g_1,$ and $\mu _2=\sqrt{\frac{\Delta _2}{%
\Delta _1}}\mu _1$. Choose $\gamma _\varphi ^{-1}=5$ $\mu $s, $\gamma
^{-1}=50$ $\mu $s, and $\kappa _{a1}^{-1}=\kappa _{a2}^{-1}=\kappa
_{b1}^{-1}=\kappa _{b2}^{-1}=20$ $\mu $s. In addition, we set $\Delta
_1/\left( 2\pi \right) =1$ GHz and $\Delta _2/\left( 2\pi \right) =0.5$ GHz.
For the parameters chosen here, the fidelity versus $b=\Delta _1/g_1$ is
plotted in Fig.~3, from which one can see that for $b=21$ a high fidelity of
$\sim 99.5\%,$ $99.5\%,$ $99.0\%,$ and $98.7\%$ can be, respectively,
achieved for the four cavities initially in the following states: (i) $%
\left| \psi ^a\left( 0\right) \right\rangle =\frac 1{\sqrt{2}}$ $\left(
\left| 00\right\rangle +\left| 11\right\rangle \right) $ and $\left| \psi
^b\left( 0\right) \right\rangle =\left| 00\right\rangle ,$ (ii) $\left| \psi
^a\left( 0\right) \right\rangle =\frac 1{\sqrt{2}}$ $\left( \left|
00\right\rangle +\left| 11\right\rangle \right) $ and $\left| \psi ^b\left(
0\right) \right\rangle =\frac 1{\sqrt{2}}$ $\left( \left| 00\right\rangle
-\left| 11\right\rangle \right) ,$ (iii) $\rho ^a\left( 0\right) =\frac
12\left( \left| 00\right\rangle \left\langle 00\right| +\left|
11\right\rangle \left\langle 11\right| \right) $ and $\left| \psi ^b\left(
0\right) \right\rangle =\left| 00\right\rangle ,$ and (iv) $\rho ^a\left(
0\right) =\frac 12\left( \left| 00\right\rangle \left\langle 00\right|
+\left| 11\right\rangle \left\langle 11\right| \right) $ and $\rho ^b\left(
0\right) =\frac 23\left| 00\right\rangle \left\langle 00\right| +\frac
13\left| 11\right\rangle \left\langle 11\right| .$

\begin{figure}[tbp]
\begin{center}
\includegraphics[bb=0 0 616 394, width=9.5 cm, clip]{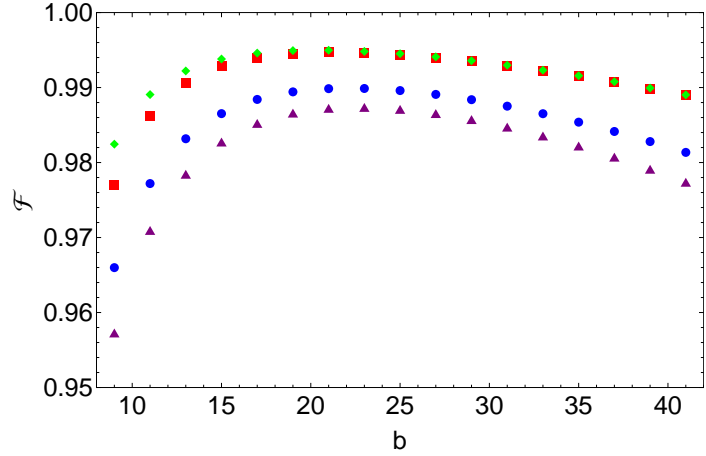} \vspace*{%
-0.08in}
\end{center}
\caption{(Color online) Fidelity versus $b$. Here, $b=\Delta_1/g_1,$ and the
blue dots, red squares, green diamonds, and purple triangles correspond to
the four initial four-cavity states (i), (ii), (iii), and (iv) described in
the text, respectively.}
\label{fig:3}
\end{figure}

For $b=21$, we have $g_{1}/\left( 2\pi \right) =\mu _{1}/\left( 2\pi \right)
\sim $ $47.6$ MHz and $g_{2}/\left( 2\pi \right) =\mu _{2}/\left( 2\pi
\right) \sim 33.7$ MHz. Note that a coupling constant $g/\left( 2\pi \right)
\sim 220$ MHz has been reported for a superconducting qubit coupled to a
one-dimensional CPW (coplanar waveguide) resonator [27], and that $T_{1}$
and $T_{2}$ can be made to be on the order of $10-100$ $\mu $s for
state-of-the-art superconducting devices at the present time [43]. For
superconducting qubits, the typical qubit transition frequency is between 4
and 10 GHz. As an example, let us consider the four cavities with frequency $%
\nu _{a1}=\nu _{b1}\sim $ 6 GHz and $\nu _{a2}=\nu _{b2}\sim $ 6.5 GHz. For
the cavity frequencies chosen above and the values of $\kappa
_{a1}^{-1},\kappa _{a2}^{-1},\kappa _{b1}^{-1},$ and $\kappa _{b2}^{-1}$
used in the numerical calculation, the required quality factors for the four
cavities are $Q_{a1}=Q_{b1}\sim 7.5\times 10^{5},$ and $Q_{a2}=Q_{b2}\sim
8.2\times 10^{5},$ respectively. Note that superconducting CPW resonators
with a (loaded) quality factor $Q\sim 10^{6}$ have been experimentally
demonstrated [44-46]. Our analysis given here demonstrates that exchange or
transfer of the states of photons in up to four cavities is feasible within
the present circuit QED technique.

Before ending this section, it should be mentioned that the impact of higher
qubit levels (above the level $\left\vert e\right\rangle $) on the fidelity
is negligible at the optimal point of Fig. 3 (where the large detuning
condition meet well). The reason for this is as follows. As long as the
large detuning condition is met, the coupler qubit $A$ remains in the ground
state and thus the excited level $\left\vert e\right\rangle $ is not
populated. Since this level $\left\vert e\right\rangle $ is not excited, all
other higher-energy levels would not be occupied during the operation. In
this sense, one can expect that the affect of the higher energy levels of
the qubit on the fidelity at the optimal point $b=21$ is negligible.

\begin{center}
\textbf{III. GENERATION OF MULTIPLE EPR PAIRS}
\end{center}

Assume that the first set of $N$ cavities is initially in the state $%
\left\vert \psi \left( 0\right) \right\rangle _{a}=\otimes
_{j=1}^{N}\left\vert 1\right\rangle _{a_{j}}$ (i.e., each cavity is in a
single-photon state) while the second set of $N$ cavities $%
(b_{1},b_{2},...,b_{N})$ is initially in the vacuume state $\left\vert \psi
\left( 0\right) \right\rangle _{b}=$ $\otimes _{j=1}^{N}\left\vert
0\right\rangle _{b_{j}}$. Based on Eq. (7), one can easily find that under
the Hamiltonian $H_{e},$ the state of the cavity system after an evolution
time $t$ is given by
\begin{eqnarray}
\left\vert \psi \left( t\right) \right\rangle _{ab}
&=&e^{-iH_{e}t}\left\vert \psi \left( 0\right) \right\rangle _{a}\left\vert
\psi \left( 0\right) \right\rangle _{b}  \notag \\
&=&\otimes _{j=1}^{N}\left[ \left( e^{-iH_{e}t}a_{j}^{+}e^{iH_{e}t}\right)
e^{-iH_{e}t}\left\vert 0\right\rangle _{a_{j}}\left\vert 0\right\rangle
_{b_{j}}\right]   \notag \\
&=&\otimes _{j=1}^{N}\left[ \cos (\lambda _{j}t)\left\vert 1\right\rangle
_{a_{j}}\left\vert 0\right\rangle _{b_{j}}\right. \ \   \notag \\
&&\left. +i\sin (\lambda _{j}t)\left\vert 0\right\rangle _{a_{j}}\left\vert
1\right\rangle _{b_{j}}\right] ,
\end{eqnarray}%
where we have used $e^{-iH_{e}t}\left\vert 0\right\rangle _{a_{j}}\left\vert
0\right\rangle _{b_{j}}=\left\vert 0\right\rangle _{a_{j}}\left\vert
0\right\rangle _{b_{j}}$ because of $\hat{a}_{j}\hat{b}_{j}^{+}\left\vert
0\right\rangle _{a_{j}}\left\vert 0\right\rangle _{b_{j}}=0$ and $\hat{a}%
_{j}^{+}\hat{b}_{j}\left\vert 0\right\rangle _{a_{j}}\left\vert
0\right\rangle _{b_{j}}=0.$

After returning to the original interaction picture by performing a
uniterary transformation $e^{-iH_{0}t}$, it is easy to find that the state
of the cavity system is given by
\begin{eqnarray}
\left\vert \psi \left( t\right) \right\rangle _{ab} &=&e^{iN\lambda
t}\otimes _{j=1}^{N}\left[ \cos (\lambda t)\left\vert 1\right\rangle
_{a_{j}}\left\vert 0\right\rangle _{b_{j}}\right.  \notag \\
&&\left. +i\sin (\lambda t)\left\vert 0\right\rangle _{a_{j}}\left\vert
1\right\rangle _{b_{j}}\right] ,
\end{eqnarray}%
where we have used $g_{j}^{2}/\Delta _{j}=\mu _{j}^{2}/\Delta _{j}=\lambda
_{j}=\lambda $ (as set above). This result (21) shows that for $\lambda
t=\pi /4,$ every two corresponding cavities $a_{j}$ and $b_{j}$ are prepared
in an entangled EPR pair of photons, i.e., $\left\vert EPR\right\rangle
_{a_{j}b_{j}}=\frac{1}{\sqrt{2}}\left( \left\vert 1\right\rangle
_{a_{j}}\left\vert 0\right\rangle _{b_{j}}+i\left\vert 0\right\rangle
_{a_{j}}\left\vert 1\right\rangle _{b_{j}}\right) $. Namely, the $N$ EPR
pairs of photons distributed in the $2N$ cavities are simulateously
generated after the operation. By performing a local operation to transfer
the state of each cavity to the qubit located in the respective cavity, the
prepared $N$ EPR pairs of photons can be tranferred onto the qubits in the $%
2N$ cavities.

\begin{center}
\textbf{IV. CONCLUSION}
\end{center}

We have proposed a method to simultaneously perform quantum state exchange
or transfer between two sets of cavities, by using only one superconducting
coupler qubit. By transferring quantum information between each qubit and
the respective cavity, the present method can be also extended to implement
quantum state exchange or transfer between two sets of $N$-qubit quantum
registers. As shown above, this work is of interest because the procedure 
for implementing quantum state exchange or transfer does not depend on the initial states of cavities
(either pure or mixed states). The quantum state exchange or transfer
between multiple pairs of cavities can be performed simultaneously, the
operation time does not increase with the number of cavities [38], and there
is no need for applying classical microwave pulses during the entire
operation so that it greatly simplifies the operation. In addition, our
analysis shows that exchanging or transferring quantum states of photons in
four ($N=2$) cavities by a coupler superconducting qubit is achieveable with
the present experimental capability. Furthermore, we have shown that this
proposal can be used to simultaneously generate multiple EPR pairs of
photons or qubits. Finally, it is noted that the superconducting coupler
qubit in this proposal can be replaced by a different type of physical
qubit, such as a quantum dot, to accomplish the same task.

\begin{center}
\textbf{ACKNOWLEDGMENTS}
\end{center}

S.H. was supported in part by DMEA. C.P.Y. was supported in part by the
National Natural Science Foundation of China under Grant No. 11074062, the
Zhejiang Natural Science Foundation under Grant No. LZ13A040002, and the funds from Hangzhou
Normal University under Grant No. HSQK0081. Q.P.S. was supported by the National Natural Science
Foundation of China under Grant No. 11147186. S.B.Z. was supported in part
by the National Fundamental Research Program Under Grant No. 2012CB921601.


\begin{references}
\bibitem{s1} J.Q. You, F. Nori, Nature (London) {\bf 474}, 589 (2011); Physics Today {\bf 58} (11), 42 (2005); Z. L. Xiang,
S. Ashhab, J. Q. You, and F. Nori, Rev. Mod. Phys. 85, 623 (2013).

\bibitem{s2} C. P. Yang, Shih-I Chu, and S. Han, Phys. Rev. A {\bf 67}, 042311 (2003).

\bibitem{s3} J. Q. You and F. Nori, Phys. Rev. B {\bf 68}, 064509 (2003).

\bibitem{s4} A. Blais {\it et al.}, Phys. Rev. A {\bf 69}, 062320 (2004).

\bibitem{s5} M. A. Sillanp\"a\"a, J. I. Park, and R. W. Simmonds, Nature (London)
{\bf 449}, 438 (2007).

\bibitem{s6} J. Majer {\it et al.}, Nature (London) {\bf 449}, 443 (2007).

\bibitem{s7} C. P. Yang, Shih-I Chu, and S. Han, Phys. Rev. Lett. {\bf 92}, 117902 (2004).

\bibitem{s8} A. Wallraff {\it et al.}, Nature (London) {\bf 431}, 162 (2004).

\bibitem{s9} I. Chiorescu {\it et al.}, Nature (London) {\bf 431}, 159 (2004).

\bibitem{s10} Y.X. Liu, L.F. Wei, and F. Nori, Europhys. Lett. {\bf 67}, 941
(2004).

\bibitem{s11} K. Moon and S.M. Girvin, Phys. Rev. Lett. {\bf 95}, 140504 (2005).

\bibitem{s12} F. Marquardt, Phys. Rev. B {\bf 76}, 205416 (2007); M. Mariantoni %
{\it et al.}, arXiv:cond-mat/0509737.

\bibitem{s13} M. Hofheinz {\it et al.}, Nature (London) {\bf 454}, 310 (2008);
H. Wang {\it et al.}, Phys. Rev. Lett. {\bf 101}, 240401 (2008).

\bibitem{s14} M. Hofheinz {\it et al.}, Nature (London) {\bf 459}, 546 (2009).

\bibitem{s15} F. Helmer and F. Marquardt, Phys. Rev. A {\bf 79}, 052328 (2009).

\bibitem{s16} L. S. Bishop {\it et al.}, New Journal of Physics {\bf
11}, 073040 (2009).

\bibitem{s17} C.P. Yang and S. Han, Phys. Rev. A {\bf 70}, 062323 (2004).

\bibitem{s18} C.P. Yang, Y.X. Liu, and F. Nori, Phys. Rev. A {\bf 81}, 062323
(2010).

\bibitem{s19} C.P. Yang, S. B. Zheng, and F. Nori, Phys. Rev. A {\bf 82}, 062326
(2010).

\bibitem{s20} P. J. Leek {\it et al.}, Phys. Rev. B {\bf 79}, 180511(R) (2009).

\bibitem{s21} A. Fedorov, L. Steffen, M. Baur, M. P. da Silva and A. Wallraff, Nature
(London) {\bf 481}, 170 (2012).

\bibitem{s22} M. D. Reed {\it et al.}, Nature (London) {\bf 482}, 382 (2012).

\bibitem{s23} M. Mariantoni {\it et al.}, Science {\bf 334}, 61 (2011).

\bibitem{s24} L. DiCarlo {\it et al.}, Nature (London) {\bf 460}, 240 (2009).

\bibitem{s25} M. Ansmann {\it et al.}, Nature (London) {\bf 461}, 504
(2009).

\bibitem{s26} J. M. Chow {\it et al.}, Phys. Rev. A {\bf 81}, 062325 (2010).

\bibitem{s27} L. DiCarlo {\it et al.}, Nature (London) {\bf 467}, 574 (2010).

\bibitem{s28} E. Lucero {\it et al.}, arXiv: 1202.5707.

\bibitem{s29} M. Mariantoni {\it et al.}, Phys. Rev. B {\bf 78}, 104508 (2008).

\bibitem{s30} F. W. Strauch {\it et al.}, Phys. Rev. Lett. {\bf 105}, 050501 (2010).

\bibitem{s31} H. Wang {\it et al.}, Phys. Rev. Lett. {\bf 106}, 060401 (2011).

\bibitem{s32} C. P. Yang, Q. P. Su, and S. Han, Phys. Rev. A {\bf 86}, 022329 (2012).

\bibitem{s33} C. P. Yang, Q. P. Su, S. B. Zheng, and S. Han, Phys. Rev. A {\bf 87}, 022320 (2013).

\bibitem{s34} J. Bylander {\it et al.}, Nat. Phys. 7, 565 (2011); H. Paik {\it et al.}, Phys.
Rev. Lett. 107, 240501 (2011); J. M. Chow {\it et al.}, ibid. 109,
060501 (2012); C. Rigetti {\it et al.}, Phys. Rev. B 86, 100506(R)
(2012); R. Barends {\it et al.}, arXiv:1304.2322.

\bibitem{s35} W. Chen, D. A. Bennett, V. Patel, and J. E. Lukens, Supercond. Sci.
Technol. {\bf 21}, 075013 (2008).

\bibitem{s36} P. J. Leek {\it et al.}, Phys. Rev. Lett. {\bf 104}, 100504 (2010).

\bibitem{s37} A. Megrant {\it et al.}, Appl. Phys. Lett. {\bf 100}, 113510 (2012).

\bibitem{s38} This is true when $N$ is less than a certain number. Beyond which, as qubit-cavity
coupling decreases, as a result of limited on-chip real estate, the time needed to complete the protocol will increase accordingly.

\end{references}
\end{document}